# Switchback Patches Evolve into Microstreams via Magnetic Relaxation

Shirsh Lata Soni[1*], Mojtaba Akhavan-Tafti[1], Gabriel Ho Hin Suen[2], Justin Kasper[1], Marco Velli[3], Rossana De Marco[4], Christopher Owen[2]

[1]Department of Climate and Space Science, University of Michigan, MI, USA
[2]Mullard Space Science Laboratory, University College London, UK
[3]Department of Earth, Planetary and Space Sciences, University of California, Los Angeles, CA, USA.
[4]National Institute for Astrophysics, Institute for Space Astrophysics and Planetology, Roma, Italy
*shirshla@umich.edu

**Abstract:**

Magnetic switchbacks are distinct magnetic structures characterized by their abrupt reversal in the radial component of the magnetic field within the pristine solar wind. Switchbacks are believed to lose magnetic energy with heliocentric distance. To investigate this switchbacks originating from similar solar source regions are identified during a radial alignment of the Parker Solar Probe (PSP; 25.8 solar radii) and Solar Orbiter (SolO; 152 solar radii). We found that 1) the dynamic and thermal pressures decrease at the switchback boundaries by up to 20% at PSP and relatively unchanged at SolO and magnetic pressure jump across the boundary remains negligible at both distances, and 2) bundles of switchbacks are often observed in switchback patches near the Sun, and in microstreams farther away. Background proton velocity ($v_p$) is 10% greater than the pristine solar wind ($v_{sw}$) in microstreams, whereas $v_p \sim v_{sw}$ in switchback patches. Microstreams contain an average of 30% fewer switchbacks than switchback patches. It is concluded that switchbacks likely relax magnetically and equilibrate their plasma with the surrounding environment with heliocentric distance. Switchback relaxation can, in turn, accelerate the surrounding plasma. Therefore, it is hypothesized that magnetic relaxation of switchbacks may cause switchback patches to evolve into microstreams with heliocentric distance. Statistical analysis of PSP and SolO switchbacks is underway to further test our hypothesis.

1. **Introduction:**

The origin and evolution of switchbacks - intense, localized rotations in the magnetic field - remain key open questions in solar wind physics (Mozer et al. 2020, Neugebauer et al. 1995; Kahler et al. 1996). These alfvénic fluctuations appear abruptly near the Sun, but link to ubiquitous turbulence farther out in the heliosphere. Unraveling this connection requires tracking switchbacks over radial distances. During the three-year span following the launch of the Parker Solar Probe (hereafter PSP) and the Solar Orbiter (hereafter SolO) missions, favorable orbital configurations have facilitated multi-point observations. These observations will provide insights into the evolution of switchbacks in relation to heliocentric distance and the solar wind conditions influenced by their progression. PSP's close-approach observations have revealed switchbacks to be ubiquitous near the Sun, occurring in both slow and fast wind streams (Kasper et al. 2019; Bale et al. 2019). Meanwhile, SolO has also detected switchback-like structures from its vantage point farther from the Sun (Fedrove et al. 2021, Horbury et al. 2021).

Switchbacks are discrete, impulsive, anti-Sunward propagating alfvenic fluctuations. Near the Sun, the amplitude of the magnetic field deflection can be larger than the field magnitude and, hence, the radial component of magnetic field (Br) can fully reverse, leading to bulk speed enhancements of up to twice the local Alfven speed, $v_A$. Therefore, switchbacks carry significant momentum and kinetic energy and appear to be an important aspect of the solar wind dynamics. Switchbacks are short on timescales of seconds to minutes, they also generally occur in patches - dense adjacent Br reversals lasting for several minutes to hours that are separated by quieter regions of near-radial magnetic field. On the other hand, microstreams are fluctuations in the solar wind speed and density associated with polarity-reversing folds in the radial component of the magnetic field that last for minutes to hours (Neugebauer et al. 1995; Neugebauer & Sterling 2021). A natural question arises here: are these magnetic structures of solar wind associated with each other?

Switchbacks evolve to release their magnetic tension and reach pressure equilibrium with their surrounding environment (Landi et al., 2005, Akhavan-Tafti et al., 2021). Akhavan-Tafti et al. (2022) revealed that rotational discontinuity (RD) type switchbacks undergo a relaxation process with an exponential decay rate of 0.06 [$Rs^{-1}$]. This process leads to the formation of magnetic discontinuities with smaller normals. The relaxation process is estimated to contribute to the transfer of up to 16% of the total reconnected magnetic energy into the surrounding plasma. Magnetohydrodynamic (MHD) simulations by Tenerani et al. (2020) show that magnetic switchbacks become increasingly unstable and eventually decay as they propagate away from the Sun.

The purpose of this study is to determine the process through which switchback magnetic energy reduces with heliocentric distance. To address this, we carefully analyze switchbacks observed from two different locations in the inner heliosphere and identify their evolution characteristics. We investigate the magnetic and plasma characteristics of isolated and adjacent switchback events observed at PSP and SolO during their radial alignment. The manuscript is organized as follows: in Section 2, we briefly introduce the data used in this study and the event identification algorithm and criteria. In Section 3, we discuss the observations and evolution characteristics of identified isolated/adjacent switchback events. Finally, in Section 4, we conclude this study and discuss future work.

## 2. Preliminary data sources and event identification algorithm

In this study, we utilize data from the PSP FIELDS instrument (Bale et al., 2016), which provides magnetic field measurements at up to 290 samples per second. Additionally, we incorporate data from the PSP SWEAP (Solar Wind Electrons Alphas and Protons) instrument suite (Kasper et al., 2016), comprising the Solar Probe Cup (SPC) and the Solar Probe Analyzers (SPAN), providing solar wind parameters at up to 4 Hz cadence. Furthermore, we include data from the SolO MAG (Horbury et al., 2020) fluxgate magnetometer, offering 8 Hz magnetic field measurements, and SolO SWA (Solar Wind Analyzer) (Owen et al., 2020), supplying electron (EAS sensor), proton, and alpha-particle (PAS sensor) 3D velocity distribution functions (VDF) with up to 4 Hz resolution.

### 2.1. Radial alignment identification:

The time intervals corresponding to the same plasma parcel observed at PSP and SolO during their radial alignment are determined using a ballistic approach. Following this method, as of the present date, we have identified 12 alignment durations for PSP and SolO. In our pursuit of pinpointing robust switchback events observed at various distances within the heliosphere, we meticulously applied three alignment selection criteria. The first criterion involves the spatial positioning of PSP in close proximity to the Sun (within < 30 Rs), while SolO is positioned at a greater distance from the Sun (> 130 Rs). The second criterion requires the period to contain the reversal of Br, and the third criterion is contingent upon the availability of high-quality magnetic and plasma measurements from both the PSP and SolO spacecraft.

Table 1 presents all 12 identified alignment durations with the corresponding distances of PSP and SolO from the Sun. By considering the first and second criteria, we narrowed down our selection to seven out of the initially reported 12 alignment durations. Subsequently, employing the third criterion, we identified only one alignment duration among the seven. The alignment duration #3 (from 2021-08-11T08:30:00 to 2021-08-12T09:30:00), when PSP was positioned at 25.8 Rs during its encounter 9 and SolO was situated at 152 Rs, satisfied all three alignment selection criteria.

Table:1 Radial alignment durations of PSP and SolO, their distance from Sun. The green shaded row indicates the identified radial alignment period.

| SN | Alignment Duration | PSP (Rs) | SolO(Rs) |
|---|---|---|---|
| 1 | 2020-09-26T20:30:00-2020-09-27T07:30:00 | 25.8 (Enc. 6) | 210 |
| 2. | 2021-04-28T20:30:00-2021-04-29T04:30:00 | 17 (Enc. 8) | 191 |
| 3. | 2021-08-11T08:30:00-2021-08-12T09:30:00 | 25.8 (Enc.9) | 152 |
| 4. | 2021-09-13T11:30:00-2021-09-24T08:30:00 | 157 | 126.8 |
| 5. | 2021-11-19T08:30:00-2021-11-20T02:30:00 | 27.9 (Enc.10) | 202 |
| 6. | 2022-02-25T12:30:00-2022-02-25T18:30:00 | 13.3 (Enc.11) | 133 |
| 7. | 2022-04-03T23:30:00- 2022-04-09T07:30:00 | 156.9 | 79.5 |
| 8. | 2022-05-30T17:30:00-2022-05-31T14:30:00 | 27.9 | 199.9 |
| 9. | 2022-09-05T23:30:00- 2022-09-06T05:30:00 | 13.3 (Enc. 13) | 150.5 |
| 10. | 2022-10-19T12:30:00-2022-10-19T13:30:00 | 163.4 | 70.9 |
| 11. | 2022-10-20T06:30:00-2022-10-23T17:30:00 | 163.4 | 73.1 |
| 12. | 2022-12-09T17:30:00-2022-12-10T09:30:00 | 25.8 (Enc. 14) | 184.9 |

Figure 1 illustrates the relative locations of the observing spacecraft in the heliosphere, and Table 2 details their specific positions within the heliosphere, with Earth's location included for reference. During the ninth encounter (E9) of PSP, it maintains a significantly close proximity to the Sun, with a heliocentric distance of 25.8 solar radii, while SolO is positioned at 152 solar radii. PSP's position is characterized by a Carrington longitude of 130 degrees and a latitude of -1.4 degrees, whereas SolO occupies a slightly different position with a Carrington longitude of 126.4 degrees and a latitude of -1.8 degrees. Additionally, both spacecraft are situated behind Earth in terms of longitudinal separation, with PSP at -85.9 degrees and SolO at -89.5 degrees. Similarly, their latitudinal separation from Earth is noteworthy, with PSP at -7.9 degrees and SolO at -8.3 degrees. The magnetic footpoint Carrington longitudes of 138.3 degrees for PSP and 170.4 degrees for SolO underscore their distinctive magnetic connections to the solar surface. These differences in location and distance, combined with their longitudinal and latitudinal separations, contribute to each spacecraft's unique vantage point for studying the evolutionary phenomena of switchbacks.

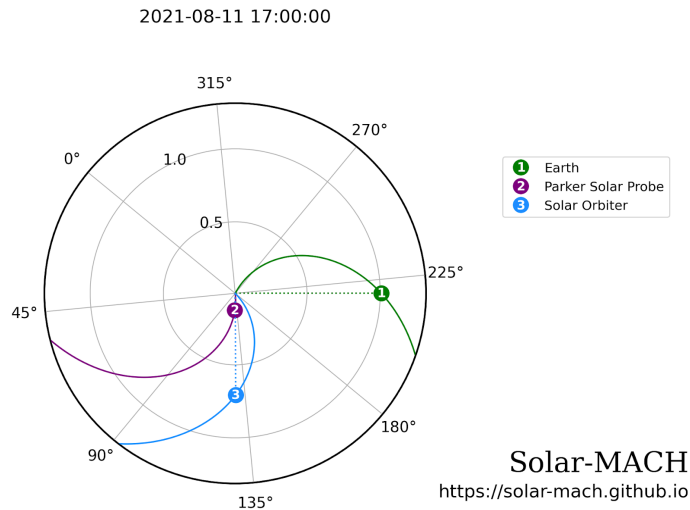

Figure 1 : Position of PSP, SolO, and Earth on August 11, 2021, at 17:00:00 UT. The grid in black corresponds to the stonyhurst coordinate systems. This polar plot is generated using the Solar-MACH tool (https://serpentine-h2020.eu/ tools/; Gieseler et al. 2022).

Table:2 Position of PSP, SolO, and Earth on August 11, 2021, at 17:00:00 UT in the inner heliosphere.

| # | Earth | PSP | SolO |
|---|---|---|---|
| Heliocent. distance [solar radii] | 215 | 28.5 | 152 |
| Carrington longitude [°] | 215.9 | 130 | 126.4 |
| Carrington latitude [°] | 6.5 | -1.4 | -1.8 |
| Longitud. separation to Earth longitude [°] | 0 | -85.9 | -89.5 |
| Latitud. separation to Earth latitude [°] | 0 | -7.9 | -8.3 |
| Magnetic footpoint Carrington longitude [°] | 278.7 | 138.3 | 170.4 |

## 2.2. Identification of robust switchback:

To identify prominent switchback candidates, the first criterion is to check the heliospheric current sheet crossing to ensure the magnetic field polarity. Then, to identify reversals in the radial component of the magnetic field (Br) in the PSP and SolO observations, we employed the well-established automated algorithm provided by Akhavan-Tafti et al. (2021) on the identified Br-reversals with clear magnetic field signatures. These signatures are defined as having five distinct regions: the leading quiet solar wind (QL), the leading transition region (TL), the spike region with a steady magnetic field (Spike), the trailing transition region (TT), and the trailing quiet solar wind (QT), along with radial velocity enhancement within the spike.

During the alignment, PSP identified a substantial negative radial magnetic field (Br) component with large variations changing to positive polarity, while SolO observed the radial component of the magnetic field changing from positive to negative polarity. The solar wind radial velocity (Vr) also exhibited significant fluctuations. In Figure 2, upper panels (a: PSP and SolO) display the pitch angle distribution (PAD) of suprathermal electrons. This population includes the 'strahl' electrons that carry heat flux away from the Sun, always directed anti-sunward along open heliospheric field lines (Feldman et al., 1975). These electrons provide information about the polarity of the magnetic field lines at the source, even if, locally, the field lines may be bent or even reversed (Owens et al., 2017). The ratio of the radial magnetic field component (Br) to the total magnetic field strength (|B|), i.e., Br/|B|, is shown in the middle panels. The radial component of velocity and local Alfven speed are plotted in panel (c). In the bottom panels (b and c), dashed blue lines indicate where Br/|B| changes polarity, but the dominating electron-PADs remain the same, representing a complete magnetic field reversal that typically characterizes the spike. We identified a total of 52 magnetic field reversals in PSP and 34 in SolO during the third alignment duration.

To identify prominent and isolated reversals, we require the ratio of the radial magnetic field component (|Br|/|B|) to shift significantly from the quiet solar wind to the spike. Specifically, we set criteria for the magnitude of the magnetic field reversal, such that |Br|/|B| should be > 0.25 within the spike and < 0.85 in the quiet period. Additionally, we look for an enhancement in velocity relative to spikes. In the case of a switchback, the radial velocity (Vr) should be greater than two times the local Alfven speed at close proximity to the Sun and greater than the local Alfven speed at further distances. Applying the above criteria, we identified one switchback candidate at each spacecraft, indicated by red dashed lines in Figure 2 (panels b and c for both PSP and SolO).

## 2.3. Identification of Switchback patches and microstreams:

During a unipolar pitch angle distribution period, a switchback patch can be defined by the following criteria: 1) a duration of minutes to hours, 2) adjacent Br reversals, 3) constant |B|, and 4) velocity enhancement within the spike regions (approximately twice to the local alfvenic velocity). On the other hand, microstreams are characterized by 1) a duration of minutes to hours, 2) adjacent Br reversals, 3) fluctuation in |B|, 4) moderate velocity enhancement (20-30 km/s from the background), and density enhancement (Neugebauer et al., 1995; Neugebauer & Sterling, 2021). In the reported alignment period, we observed four switchback patches at PSP (orange shaded regions in Figure 2, b and c) and three microstreams at SolO (yellow shaded regions in Figure 2, b and c). It is important to note that switchback patches are only observed near the Sun (at PSP), while microstreams are solely observed farther away (at SolO).

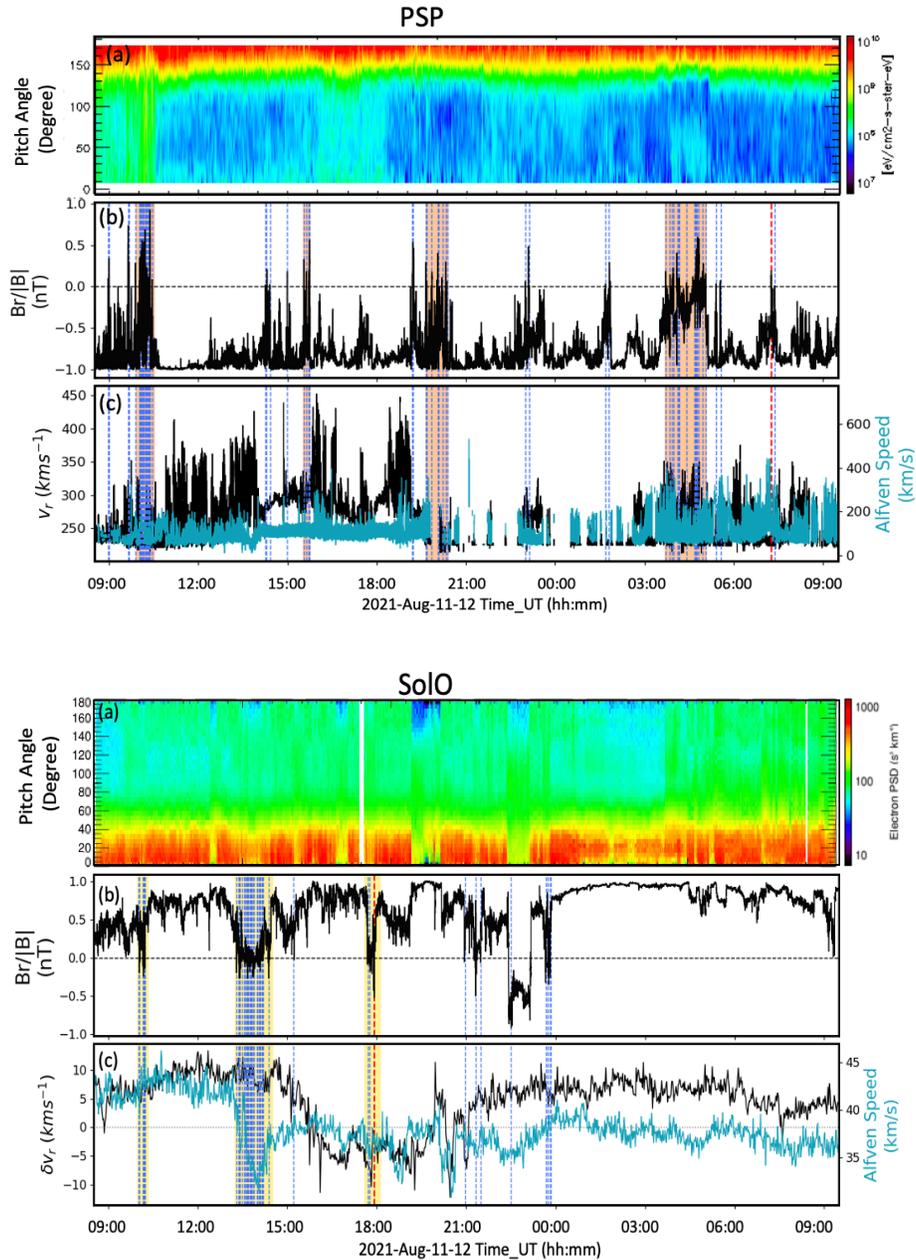

Figure 2. Shows the time interval from 2021-08-11, 08:30:00 to 2021-08-12 09:30:00, when PSP and SolO were radially aligned. In both PSP and SolO, the top panels (a) show the normalized pitch angle distributions of suprathermal electrons (e-PADs) at an energy of 314 eV, middle panels (b) present the radial to total magnetic field ratio (Br/|B|). Bottom panels (c) show the radial component of velocity along with local alfven speed. Dashed vertical blue lines indicate the number of magnetic field reversals during the spacecraft alignment period and the red dashed lines are for the reported events for PSP and SolO. Orange shaded regions in PSP present the switchback patches and yellow shaded regions in SolO indicate the microstreams.

## 3. Results:
### 3.1. Evolution of switchbacks observed at PSP and SolO

In this section, we identify the observational signature of candidate switchbacks observed on 12 August 2021 at 07:00:00 UT at PSP and on 11 August 2021 at 17:55:00 UT at SolO (Figure 3). In both plots, panels (a) and (b) display the magnetic field B with Br and proton bulk velocity v in the RTN-coordinates. To ease visualization, we subtract the average proton bulk velocity <v> across the sampled interval from the data. The background solar wind speed was 232 km/s for the observed switchback at PSP, while it was 324 km/s for the switchback observed at SolO. Panels (c) show the plasma density and temperature. Panel (d) presents the magnetic pressure (Pmag), and the bottom panel (e) displays the dynamic (Pdyn) and thermal (Pth) pressure .

PSP identified a substantial negative radial magnetic field (Br) component with significant polarity changes, transitioning from negative to positive, while SolO observed fluctuations shifting to negative polarity from positive. The solar wind radial velocity (Vr) is also highly variable and dominant compared to Vt and Vn. The variability in flow speed exhibited in switchbacks is directly related to the magnetic field (Raouafi et al., 2023). In the reported event, the jump in velocity is higher (7.26%) compared to SolO (1.37%). In PSP's switchback, plasma density sharply decreases in the spike region, by approximately -25% (200 $cm^{-3}$ magnitude), compared to the quiet solar wind. In the SolO switchback, a slight decrease (-2.96%) is observed in spikes compared to the leading quiet and transition regions. While plasma temperature in both PSP and SolO doesn't show any significant variation in different regions of switchbacks. The analysis shows that Pmag is relatively unchanged in PSP and SolO observations. In contrast, Pdyn and Pth decrease at the switchback boundaries by up to -20% at PSP, while they remain relatively unchanged at SolO.

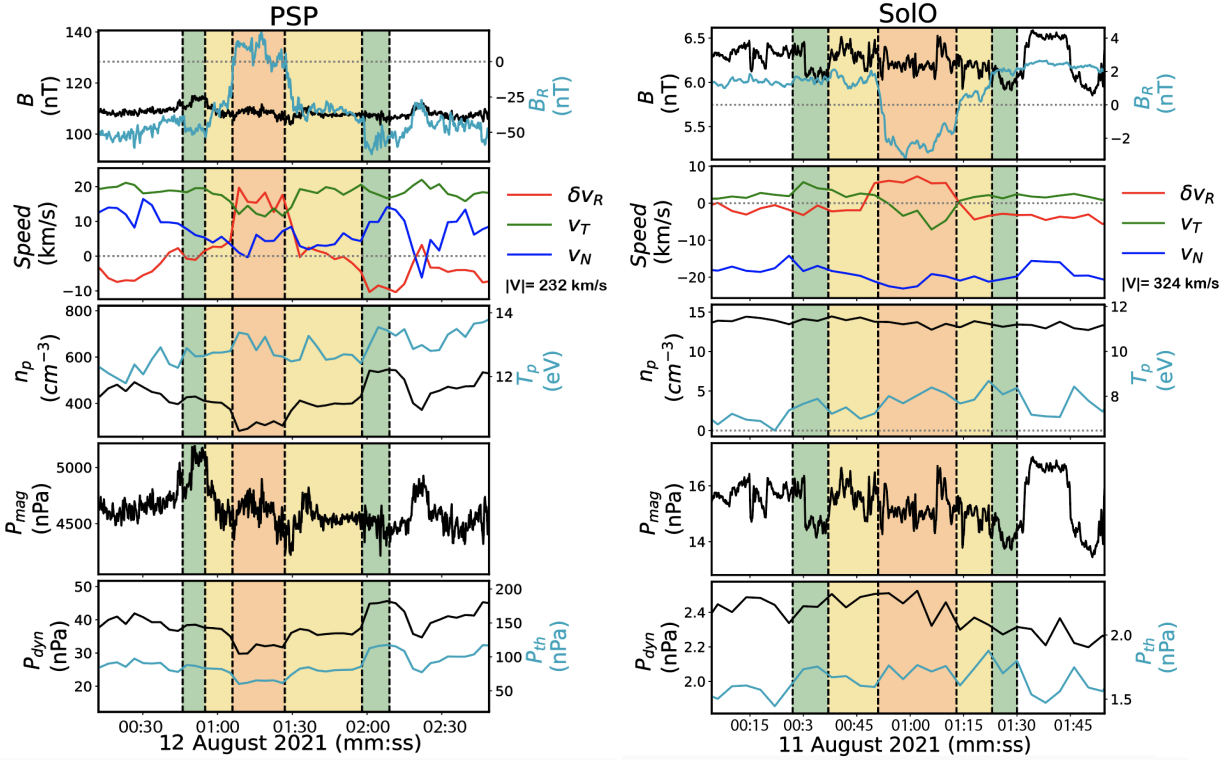

Figure 3: (Left: PSP and Right : SolO), Panels top to bottom: (a) Magnetics field (|B|, Br), (b) Velocity plasma moment (Vrtn), (c) plasma density and temperature (d) magnetic pressure. (e) dynamic and thermal pressure. Green shaded regions represent the QL and QT respectively, yellow shaded regions represent the TL and TT, and orange shaded regions represent the Spike.

In solar wind, alpha-proton signature is largely governed by conditions at the source region and thus, provides useful information about its properties and the release mechanisms involved in the formation of the solar wind ( Bochsler 2007; Aellig et al. 2001; Kasper et al. 2012; Huang et al. 2016b, 2023; Fu et al. 2018). Figure 4 shows the time series of various proton and alpha population parameters measured in the reported switchbacks. Panel a) shows the magnetic field B, b) shows the helium abundance ratio $A_{He} = (n_\alpha/n_p)*100\%$, where $n_\alpha$ and $n_p$ are the alpha and proton number density respectively. Panels c) and d) show the parallel ($T_\parallel$, solid line) and perpendicular ($T_\perp$, dotted line) temperatures of the proton core $T_c$ and alphas $T_\alpha$, respectively. Panel e) shows the signed magnitude of the alpha-proton velocity difference vector, $\Delta v_{\alpha p} = |v_\alpha - v_p|*sgn(v_{\alpha,R} - v_{p,R})$, normalized to the local Alfven speed $v_A$ (Reisenfeld et al. 2001; Ďurovcová et al. 2017; Fedorov et al. 2021). For the SolO event, we obtained this data using the methods developed by De Marco et al. (2023).

Both of the observed switchbacks are embedded within slow solar wind streams containing helium-poor ($A_{He} \leq 1\%$) plasma. Inside the spike region of both events, $A_{He}$ decreases slightly compared to its value in the surrounding solar wind. At PSP, $A_{He}$ decreases from ~0.2% in the TL and TT regions to 0.1% inside the spike region. $A_{He}$ in the TL and TT regions does not change compared to the QL, QT, and switchback exterior regions. In the case of the SolO switchback, $A_{He}$ is higher before and during the switchback encounter compared to after the encounter, and is overall smaller compared to PSP. $A_{He}$ decreases to 1% compared to the QT region. It indicates that reported switchbacks observed within slow solar wind having decreased $A_{He}$ (helium-poor population) should be threaded to helmet streamer regions (Kasper et al. 2007, 2012; Alterman et al. 2018; Alterman & Kasper 2019).

In slow solar wind the parallel temperature of proton/alpha should be higher than the parallel proton/alpha temperature (Li et al., 2023). However, there is no significant variation in either component that can be attributed to the switchbacks. $\Delta v_{\alpha p}$ increases toward the Sun but the magnitude is mainly below $v_A$, and alpha particles usually move faster than protons near the Sun, based on PSP observations. However, the high $\Delta v_{\alpha p}/v_A$ in the solar wind may be associated with the very low local Alfven speed when magnetic field lines change polarity, or it may point to the preferential acceleration of alpha particles (Isenberg & Hollweg 1983; Kasper et al. 2017). The average $\Delta v_{\alpha p}/v_A$ is observed between -0.15 and -0.2 outside the spike region although there are large fluctuations, particularly in the TL and QT regions. Inside the spike region, we observe a minimum average $\Delta v_{\alpha p}/v_A$ of ~-0.4. The negative values of $\Delta v_{\alpha p}/v_A$ could be a result of waves that slow down alpha particles but accelerate the protons as the energy of alpha particles exceeds that of protons (Durovcova et al. 2017).

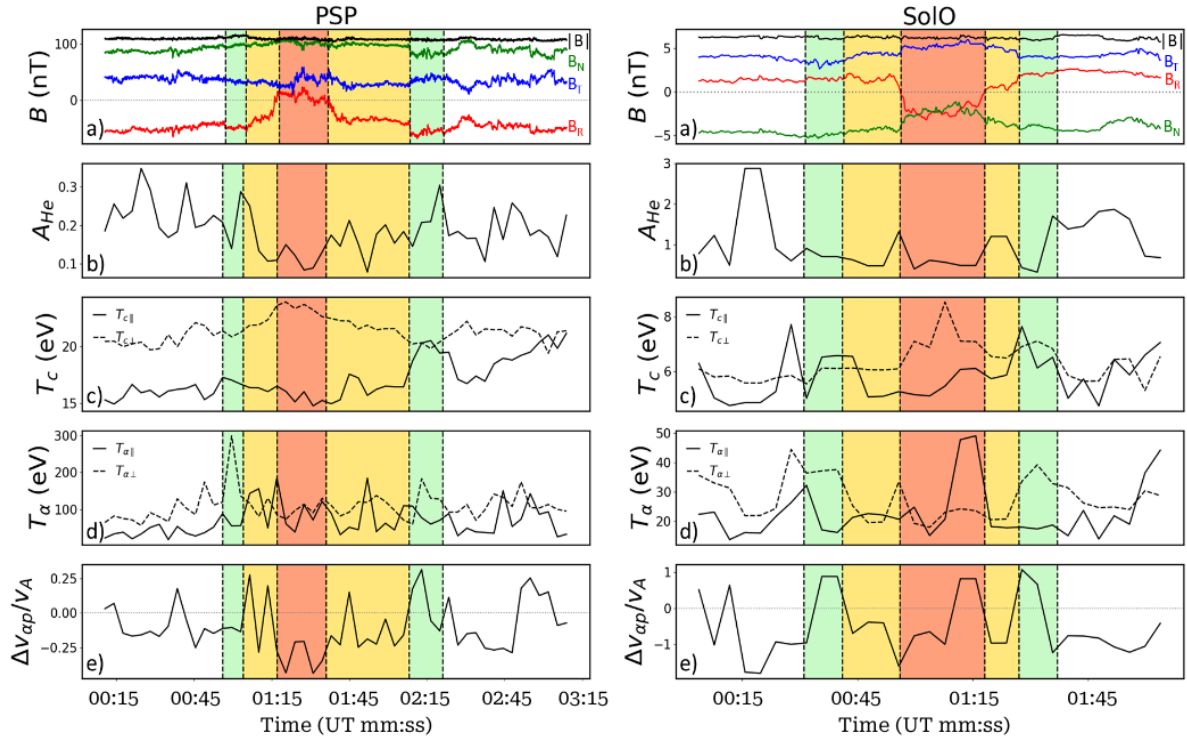

Figure 4: (Left: PSP and Right : SolO), Panels top to bottom: (a) Magnetics field (|B|, Brtn), (b) He abundance, (c) Proton core temperature (Tperp, Tpar), (d) Alpha temperature (Tperp, Tpar), (e) Alpha-Proton differential Speed. Green shaded regions represent the QL and QT respectively. Yellow shaded regions represent the TL and TT and orange shaded regions represent the spike.

Table: 3 Estimated jump (QL to spike) of magnetic field and plasma parameters for candidate switchbacks.

| Jump (QL to Spike) | PSP (25.8 Rs) | SolO (152 Rs) |
|---|---|---|
| **Duration of Switchbacks** | 85 Sec | 72 Sec |
| **|B|** ($1/r^2$) (nT) | -4.06% | 0.48% |
| **Density** ($1/r^2$) (cm^-3) | -25.14% | -2.96% |
| **Temp** ($1/r^{y-1}$) (eV) | 1.03% | 7.35% |
| **|V|** (km/s) | 7.26% | 1.37% |
| **P_Mag** ($B^2/2\mu 0$) (nPa) | -0.82% | 0.63% |
| **P_Dyn** ($m_p N_p V_p^2$) (nPa) | -19.79% | -2.84% |
| **P_Th** ($N_p KT_p$) (nPa) | -25.3% | 5.76% |

### 3.2. Evolution of Switchback bundles observed at PSP and SolO

Figure 5 shows an example of a switchback patch observed at PSP and a microstream observed at SolO, where the top panel shows the radial component of the magnetic field Br to the magnitude B averaged followed by velocity components, density temperature, magnetic, thermal and dynamic pressure. At PSP the shaded region indicates the switchback patch for the duration of 80 min contains 18 Br reversals associated with velocity jump along with constant total magnetic field. However at SolO, the shaded region depicts the boundaries of a microstream for a duration of ~20 min with a slight enhancement in velocity and density. Notably, in the microstream, only 6 Br reversals were found.

By analyzing all identified switchback patches at PSP and microstreams at SolO, we observed that the microstreams contain a 30% lesser number of switchbacks to the switchback patches. Additionally, we found that, in microstreams, the background proton velocity (vp) is approximately 10% higher than the pristine solar wind, while in switchback patches, vp is comparable to the solar wind velocity (vsw). This underscores that the microstreams could potentially emerge as a product of continuous and lasting increases in velocity that stem from a sequence of switchbacks.

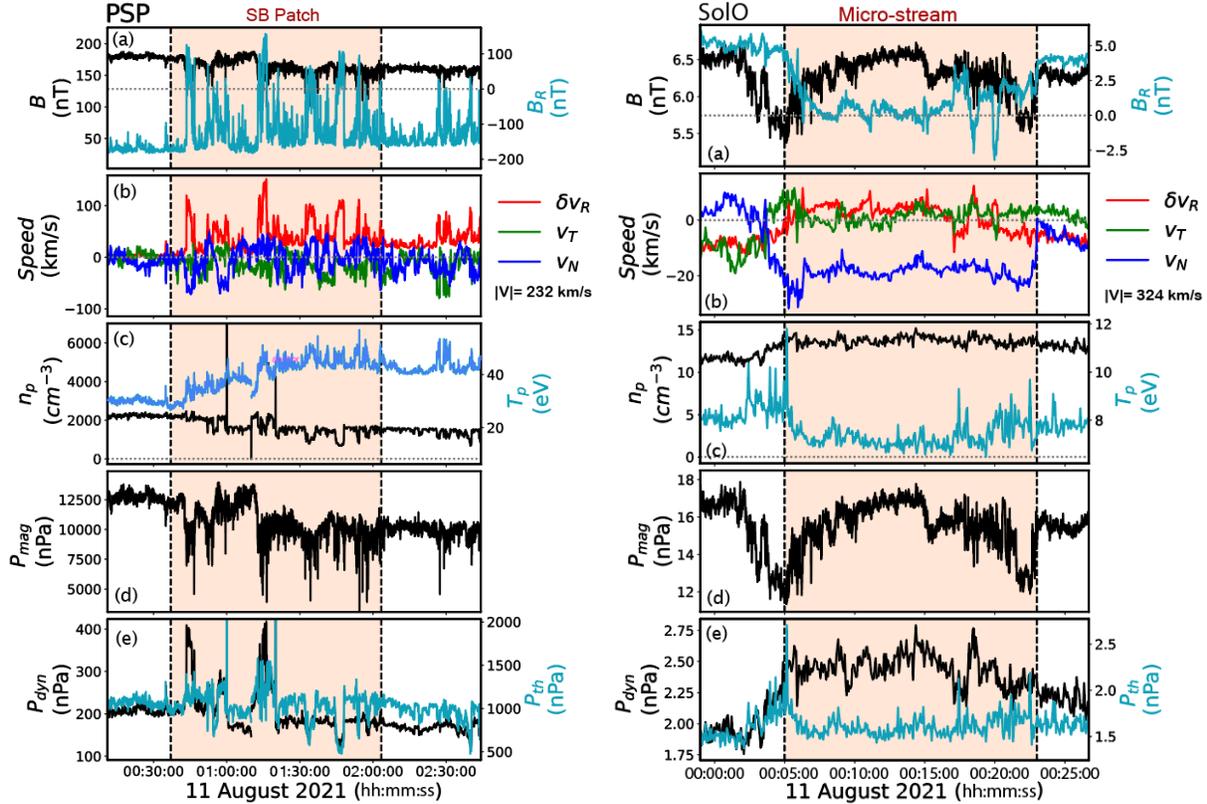

Figure 5: PSP (left) and SolO (right): Panels top to bottom: (a) magnetic field (|B|, $B_r$), (b) velocity plasma moment ($V_{rtn}$), (c) proton density and temperature, (d) magnetic pressure, (e) dynamic and thermal pressure. Orange shaded region is switchback patch at PSP and microstream at SolO.

## 4. Discussion and conclusion:

This investigation aims to compare switchback properties at two different heliocentric distances during a radial alignment of PSP and SOlO, with the goal of studying the evolution characteristics of switchbacks. The key results, based on the observed magnetic and plasma characteristics of switchbacks observed at PSP and SolO, are as follow:

1. In the slow solar wind, switchback events are observed with helium-poor plasma ($A_{He}$ < 1%) and low $\Delta v_{\alpha p}/v_A$ population, indicating that switchbacks may have originated from a similar source region (helmet or streamer-like structures) and possibly generated via magnetic reconnection (Durovcova et al. 2017, 2019; Fu et al. 2018).

2. Magnetic and plasma observations at the switchback leading transition regions show that $P_{mag}$ is relatively unchanged in both PSP and SolO observations. In contrast, $P_{dyn}$ and $P_{th}$ sharply drop across the switchback boundary at PSP by up to ~-20%, while they remain relatively unchanged at SolO.

    a. The magnetic and plasma observations at the switchback boundary regions at PSP and SolO indicate that switchback plasma population may come to equilibrium with the surrounding environment with heliocentric distance. The equilibrium further points to a possible flow of plasma and energy across switchback boundaries with heliocentric distance, in agreement with Akhavan-Tafti (2022) who argued that switchbacks are dominantly permeable, rotational-type magnetic discontinuities.

    b. The rotational discontinuity type boundaries near the Sun likely influence the dynamics of the switchbacks (Akhvan-Tafti et al. 2022), enabling the efficient exchange of material and energy that, in turn, contributes to the overall stability and equilibrium observed across varying distances within the heliosphere. .

3. Observation shows that bundles of switchbacks are observed as switchback patches at PSP and  solely observed as microstreams at SolO. Observed microstreams contain 30% fewer switchbacks than switchback patches. Furthermore, notable variations emerge in the background proton velocity ($v_p$) within these phenomena. In microstreams, the background proton velocity ($v_p$) is approximately 10% greater than the pristine solar wind, while in switchback patches, $v_p$ is approximately equal to the solar wind velocity ($v_{sw}$). Additionally, switchback dynamic pressure jump is greater in switchback patches than microstreams.

    a. Lesser number of switchbacks observed in microstreams suggest that some of the switchbacks inside the switchback patches dissipate with distance.

    b. The dissipation of switchbacks can result in accelerating background plasma. The microstreams peak might have been a consequence of a relaxation of adjacent switchbacks with heliocentric distance.

    c. In this scenario, it is conceivable that the switchback population magnetically equilibrates with the surrounding environment as it moves further in the heliosphere by dissipating its energy to the background. This, in turn, can result in switchback patches that may evolve into microstreamers over heliocentric distance.

In summary, Figure 6 presents our proposed mechanism for how switchbacks' magnetic energy reduces with heliocentric distance. Switchback patches are noted to occur more frequently in proximity to the Sun, while microstreams are more commonly observed at greater distances. Within switchback patches, a notable feature is the relatively higher concentration of switchbacks compared to microstreams. Our studies suggest that the accumulated and persistent velocity enhancement in the micro streams observed at SolO stemmed from the relaxation of a series of switchbacks observed near the Sun.

Furthermore, a discernible difference in the dynamic pressure jump associated with switchbacks in patches versus micro streams is observed. The dynamic pressure jump is found to be greater (approximately 20%) in the switchback population, indicating a more substantial impact or influence of these magnetic structures when they occur near the Sun. One intriguing aspect of switchback behavior is the magnetic relaxation process that takes place as they move farther away from the Sun. This relaxation allows the flow of solar wind across the boundaries of the switchbacks. Consequently, resulting in the observed lower curvature and smaller $\nabla P_{dyn}$ further away from the Sun.

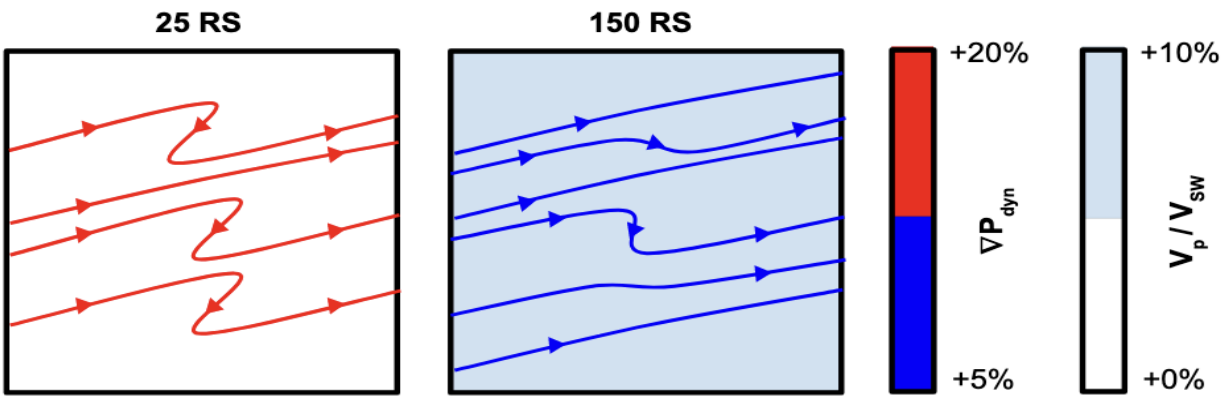

Figure 6. Concept illustration of the spatial and temporal evolution of a magnetic switchback. The color bars indicate dynamic pressure ($P$dyn) and relative velocity (Vp/Vsw).

The combined measurements from PSP and SolO significantly advance our understanding of how switchbacks evolve throughout their propagation, unveiling the remarkable endurance of these transients. Future work will focus on the statistical analysis of switchbacks identified at various distances in the heliosphere to validate their origins, formation, and evolution. Source region mapping will be crucial for understanding 'in-situ' and/or 'ex-situ' generation mechanisms of switchbacks. Additionally will provide key insights into the nature of magnetic and plasma interactions shaping the inner heliosphere, as well as reveal their contribution to solar wind heating.

5. Acknowledge

The data used in this paper can be downloaded from spdf.gsfc.nasa.gov. The authors are grateful for the dedicated efforts of the PSP and SolO team. This work was supported by NASA contract Nos. NNN06AA01C, 80NSSC20K1847, 80NSSC20K1014, and 80NSSC21K1662.